\documentclass{eg/egpubl}

\usepackage{eg/eg2026}

\ShortPresentation  

\usepackage[T1]{fontenc}
\usepackage{eg/dfadobe}  

\BibtexOrBiblatex
\usepackage[backend=biber,bibstyle=eg/EG,citestyle=alphabetic,backref=true,mincitenames=1,maxcitenames=2,natbib=true]{biblatex} 
\renewcommand*{\mkbibnamefamily}[1]{#1}
\renewcommand*{\mkbibnamegiven}[1]{#1}
\electronicVersion
\PrintedOrElectronic

\ifpdf \usepackage[pdftex]{graphicx} \pdfcompresslevel=9
\else \usepackage[dvips]{graphicx} \fi

\usepackage{eg/egweblnk} 

\usepackage{caption,cuted,amssymb, amsmath}
\usepackage{algorithm}
\usepackage{algorithmic}
\usepackage{booktabs}
\usepackage{multirow}
\usepackage{xspace}
\usepackage{cleveref}
\usepackage{hyperref}

\newcommand{\mypara}[1]{\noindent\textbf{#1}}
\newcommand{\visacd}{VisACD\xspace}

\title{VisACD: Visibility-Based GPU-Accelerated\\Approximate Convex Decomposition \vspace{-0.5cm}}

\author[Egor Fokin \& Manolis Savva]{
    \parbox{\textwidth}{\centering Egor Fokin$^{1}$ \quad Manolis Savva$^{1}$}
    \\
    {\parbox{\textwidth}{\centering $^1$Simon Fraser University} 
    }
    \\
    {\parbox{\textwidth}{\centering \href{https://3dlg-hcvc.github.io/visacd/}{3dlg-hcvc.github.io/visacd}} \vspace{-0.5cm}
    }
}

\begin{document}

\teaser{
  \includegraphics[width=\textwidth]{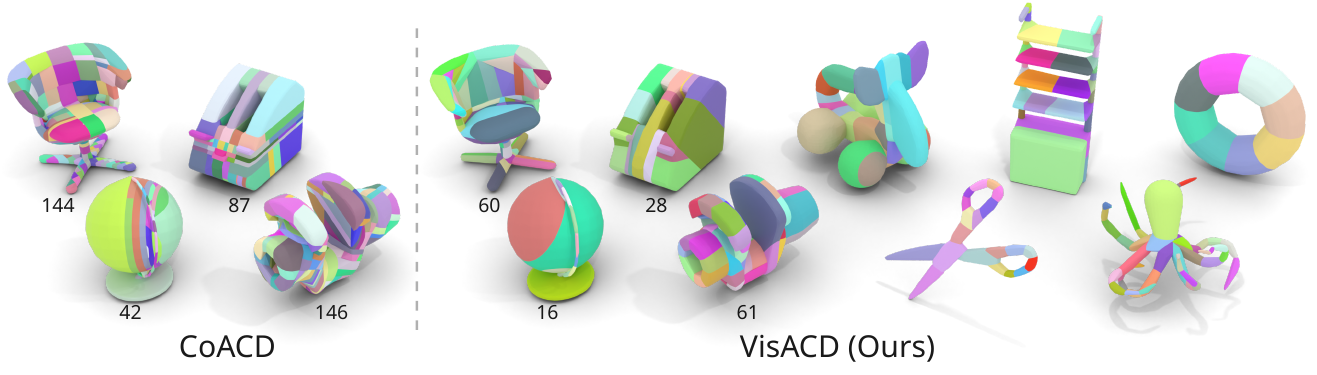}
  \caption{We present \visacd, a visibility-based GPU-accelerated intersection-free approximate convex decomposition (ACD) algorithm. \visacd is rotation-equivariant and thus not sensitive to input mesh orientation. Our experiments show that \visacd outperforms prior methods producing accurate decompositions with fewer parts (indicated by numbers).}
  \label{fig:teaser}
}

\maketitle

\begin{abstract}
Physics-based simulation involves trade-offs between performance and accuracy.
In collision detection, one trade-off is the granularity of collider geometry.
Primitive-based colliders such as bounding boxes are efficient, while using the original mesh is more accurate but often computationally expensive.
Approximate Convex Decomposition (ACD) methods strive for a balance of efficiency and accuracy.
Prior works can produce high-quality decompositions but require large numbers of convex parts and are sensitive to the orientation of the input mesh.
We address these weaknesses with \visacd, a visibility-based, rotation-equivariant, and intersection-free ACD algorithm with GPU acceleration.
Our approach produces high-quality decompositions with fewer convex parts, is not sensitive to shape orientation, and is more efficient than prior work.



\end{abstract}
\section{Introduction}

Representing shapes as a set of convex hulls is an acceleration technique used widely in physics and game engines.
This representation enables efficient algorithms such as checking whether a point lies inside the mesh \cite{snoeyink2017point}, computing distances between two objects \cite{gilbert2002fast} or checking whether two meshes intersect \cite{bergen1999fast}.
In practice, sets of convex hull `colliders' are authored by content creators through fitting primitives such as boxes or capsules to a mesh asset.
This is a time-consuming process, and the resulting colliders can differ substantially from the original mesh asset.
Multiple algorithms were created to automate this process and produce more accurate convex decompositions.

Exact convex decomposition \cite{chazelle1981convex} attempts to decompose a shape into convex hulls such that the shape of the decomposition is exactly the same as the input mesh.
Such algorithms are slow and produce thousands of parts negating the intended efficiency gains.
A different approach is Approximate Convex Decomposition (ACD) \cite{10.1145/997817.997889}, which decomposes objects into parts that approximately match the initial shape.
This relaxed condition allows algorithms that are faster and produce decompositions with orders of magnitude fewer parts.

In this paper, we present \textbf{a)} \visacd, a method that produces convex decompositions closer to the initial mesh and with lower part numbers, that is not sensitive to mesh orientation and that is more efficient, compared to baselines from prior work. \textbf{b)} A concavity metric tailored for efficient cutting plane computation. \textbf{c)} A parallelization of our algorithm using NVidia OptiX and CUDA.


\section{Related Work}

\mypara{Applications for ACD methods.}
Recent works \cite{li2024evaluating, gu2023maniskill2} use ACD to create colliders for different objects in robot grasping simulations. NERF-Texture \cite{huang2023nerf} uses CoACD as volume-preserving smoothing.
V-HACD \cite{mamou2016volumetric} was adapted by Unreal Engine 4 for automatic collider creation of 3D objects.
The quality of the decompositions directly impacts the accuracy and efficiency in the mentioned applications.

\mypara{Concavity measures for ACD.}
ACD algorithms use a concavity measure that they minimize and use in evaluation.
Some methods use surface-based metrics to measure concavity.
These metrics include distances between the mesh surface and its convex hull \cite{10.1145/997817.997889, GHOSH2013494}, or between the shortest geodesic path on the mesh and the convex hull \cite{liu2016nearly}.
HACD \cite{5414068} projects the vertices onto the convex hull and measures the distances between the points and their projections.
\citet{wei2022approximate} show that these metrics fail to capture differences of volume properties.
Other methods use volume-based metrics, such as volume ratio between a mesh and its hull \cite{attene2008hierarchical, mamou2016volumetric, thul2018approximate}, and CoACD \cite{wei2022approximate} uses a mixture of volume-based and surface-based metrics.
Another family of metrics uses pairwise visibility of surface points (two points are considered visible to each other if the segment between them lies fully outside the mesh).
Some works \cite{kaick2014shape, asafi2013weak} use a percentage of pairs that are mutually visible.
Other works \cite{5540225, 6126256} calculate the distance between those connections and the surface of the mesh.
This metric can be easily computed and has useful properties that we discuss later.
Our method uses an efficient visibility-based metric at its core, computing the total combined length of all segments between mutually visible vertices.

\mypara{Prior work on ACD algorithms.}
\citet{10.1145/997817.997889} introduced a surface-based approach that cuts meshes at regions of high concavity, with later works improving concavity measures \cite{GHOSH2013494}, cut paths \cite{liu2016nearly}, or using triangle merging \cite{5414068,kucskonmaz2024surface}.
However, surface-based methods struggle to preserve volume, especially in meshes with holes or cavities.
To address this, volumetric ACD methods were proposed, including tetrahedral merging \cite{attene2008hierarchical}, point-patch clustering \cite{kaick2014shape}, spectral clustering \cite{asafi2013weak}, and Reeb-graph-based approaches \cite{5540225,6126256}.
While these methods better preserve volume, they do not prevent convex hull intersections, limiting their applicability in animation and simulation.
More recent works utilize cutting planes as a main tool of decomposition to prevent convex hull intersections.
V-HACD \cite{5414068} uses axis-aligned cutting planes to decompose a voxelized representation of the mesh.
CoACD \cite{wei2022approximate} improves on this by using an MCTS algorithm instead of a greedy one and by employing a regular mesh representation together with a volumetric concavity measure to preserve volume properties.
\citet{thul2018approximate, andrews2024navigation} additionally sample planes that contain edges of maximum concavity.
Such methods evaluate the cutting planes by performing the cut and computing the concavity of resulting pieces, which is highly inefficient, limiting the number of planes that can be sampled at every iteration.

\begin{figure}
    \centering
    \includegraphics[width=0.7\linewidth] {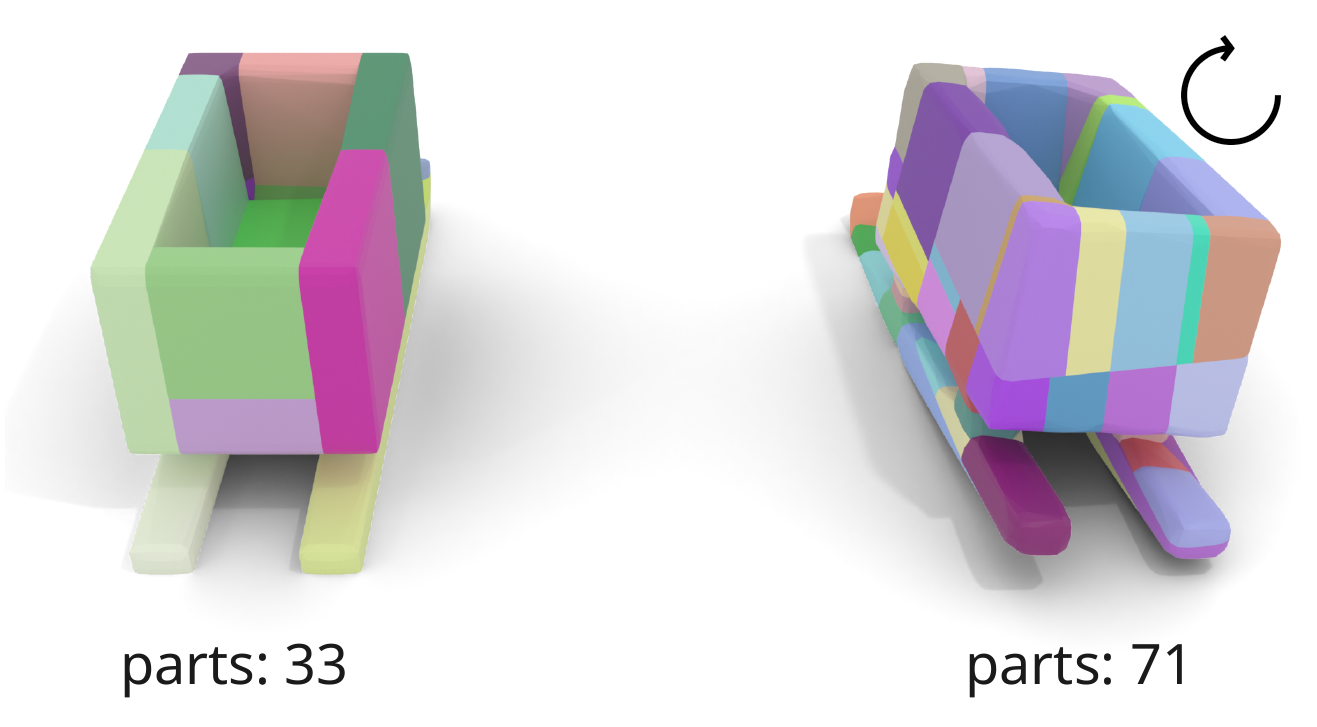}
    \caption{Sensitivity of axis-aligned plane methods to orientation.}
    \label{fig:coacd_rotation}
\end{figure}

Our approach also uses cutting planes.
However, by using visibility-based concavity we can compute a plane \textit{value} without cutting the mesh, which allows us to sample thousands of planes per iteration and achieve better results.
Our plane sampling strategy also makes the method rotation-equivariant, eliminating sensitivity to mesh orientation in axis-aligned plane methods (\Cref{fig:coacd_rotation}).

\section{Method}

\subsection{Visibility-Based Concavity Metric}

We use the definition of weak convexity proposed by \citet{asafi2013weak} and define \textbf{visibility edges} $e_i$ as edges between two mesh vertices that lie outside of the mesh and do not intersect it.
An important property is that a convex mesh has zero visibility edges and the more concavities there are in a mesh, the more visibility edges.

One way to measure the concavity of the mesh is to use the number of such edges.
However, not all visibility edges are equally important.
A better way to compute the cost of a visibility edge was proposed by \citet{6126256}.
They take the maximum perpendicular distance between the edge and the mesh surface.
Unfortunately, this is expensive to compute, especially in 3D.
Instead, we use the length of an edge as the cost.
This is motivated by the observation that edges farther away from the surface tend to have higher length.
We compute the concavity of the mesh $C^*$ using a combined length of all visibility edges $e_i$:
$C^*(M) = \sum_{i} \|e_i\|_2$.
We also define the concavity of the decomposition $D=\{M_1, M_2, ..., M_n\}$ as:
$C^*(D) = \sum_{M\in D}C^*(M)$.
Because this concavity measure is based on mesh vertices rather than the volume enclosed by the mesh, it depends on mesh topology and cannot be used to compare decompositions across different meshes.
Therefore, for evaluation purposes we use the \textit{collision-aware concavity measure} $C$ by \citet{wei2022approximate}.
This metric takes the minimum between Hausdorff distance and scaled volume difference.

\begin{figure}
    \centering
    \includegraphics[width=1\linewidth,
    trim=0 3cm 0 2cm,
    clip
  ] {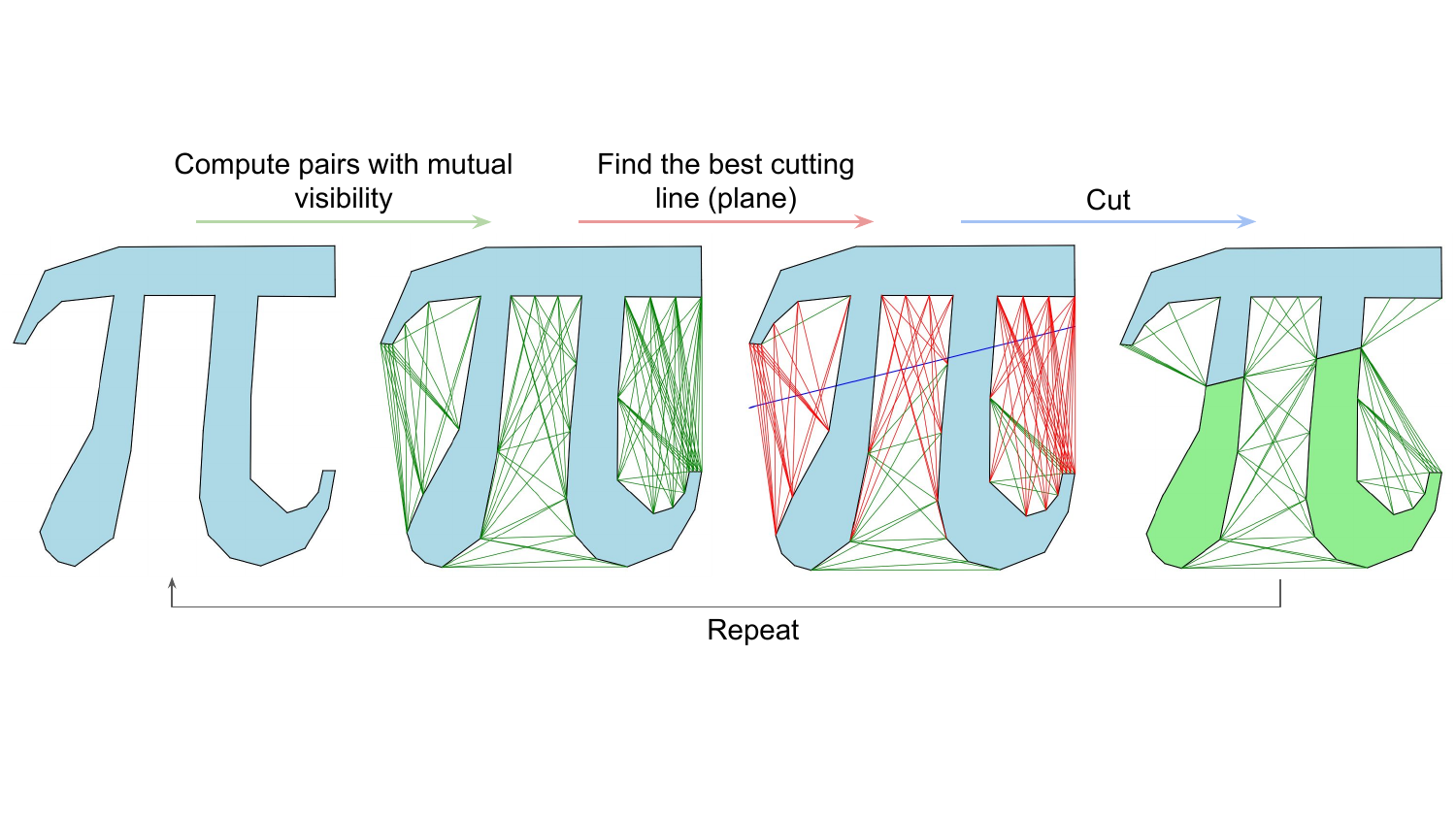}
    \caption{We use pairwise point visibility to compute the \textbf{visibility edges} and approximate the concavity of the mesh. We sample $k$ visibility edges and compute bisecting orthogonal planes for each. The value of the plane $Q_p(M,E)$ is the sum of cut visibility edge lengths. We select the plane with highest value to make the cut.}
    \label{fig:method}
\end{figure}

\subsection{Visibility Edge Computation}

Our main insight is that ``point visibility'' algorithms are highly parallelizable. Prior work does not leverage this and instead extrapolates from a subset of visibility edges \cite{kaick2014shape} or uses visibility between Reeb graph vertices \cite{5540225, 6126256}
We implement a parallel GPU-accelerated algorithm to compute \textit{all} visibility edges, leading to high precision without increasing computation time.

To compute the visibility edges of a mesh \(M\), we first construct a cage mesh \(M_c\) offset by \(\varepsilon = 0.03\) from the surface of \(M\).  
An edge \(e\) is classified as a \emph{visibility edge} if and only if  
$\left(e \cap M_c \neq \varnothing\right)
\;\land\;
\left(e \cap M = \varnothing\right)$.
In practice, these intersection tests are efficiently evaluated using ray–mesh queries implemented in NVIDIA OptiX.
This method of computing visibility edges also filters out edges that are too close to the mesh and do not significantly impact the concavity.
By changing \(\varepsilon\), one can manipulate how many such edges are filtered out.

\subsection{Plane Value}

Given a mesh $M$, a cutting plane $p$ splits it into two meshes $M_1$ and $M_2$.
We want to find a plane $p$ that minimizes $C^*(\{M_1,M_2\})$.
Let $I_P(e_i) \in \{0,1\}$ be the indicator that $p$ cuts the visibility edge $e_i$ and $e'_i$ be the new visibility edges created after the cut.
Then:
\begin{equation}
C^*(\{M_1,M_2\}) = C^*(M) - \sum_{i}I_P(e_i)||e_i||_2 +  \sum_{i}||e'_i||_2
\label{eq:cost1}
\end{equation}
The combined length of the newly created edges $\sum_{e'_i}||e'_i||_2$ is complex to compute and, for high vertex densities, contributes little to the overall concavity. Therefore, we can simplify:
\begin{equation}
C^*(\{M_1,M_2\}) \approx C^*(M) - \sum_{i}I_P(e_i)||e_i||_2
\label{eq:cost2}
\end{equation}
Then:
\[
\begin{aligned}
\arg\min_{p}\, C^{*}(\{M_1, M_2\})
&\approx 
\arg\min_{p}\!\left( C^{*}(M) - \sum_{i} I_{p}(e_i)\,\|e_i\|_2 \right) \\
&= 
\arg\max_{p} \sum_{i} I_{p}(e_i)\,\|e_i\|_2 .
\end{aligned}
\]
From this, we get the value function $Q_p(M,E)$ of a plane $p$:
\begin{equation}
Q_p(M,E) = \sum_{i} I_{p}(e_i)\,\|e_i\|_2
\label{eq:value}
\end{equation}
This value is interpretable as the combined length of all visibility edges that the plane $p$ cuts. 

\subsection{Candidate Planes}

Choosing plane candidates is a crucial step in ACD algorithms. CoACD~\cite{wei2022approximate} and V-HACD~\cite{mamou2016volumetric} consider only axis-aligned cutting planes spaced by small epsilon values and \citet{thul2018approximate} uses planes that intersect \textit{concave} edges.
The reason behind these choices is that the number of planes they can check at each step is quite low ($60$ for CoACD), which motivates setting strict limitations on the planes sampled.
Our efficient value function enables sampling many more planes (1000+) at each step.
Similarly to our concavity measure and value function, we sample cutting planes based on visibility edges.
We pick $k$ random edges and assign a plane to each of them such that each plane is orthogonal to the corresponding edge and bisects it.
Additionally, the planes that correspond to the largest flat surfaces of the mesh are more likely to be cutting planes.
For this reason, we extract these planes, add them to the set of candidates, and double their value.

\subsection{Greedy Algorithm}

To arrive at the final decomposition $D$ from the initial mesh $M$, we follow a simple greedy algorithm (\Cref{fig:method}).
At each step, we pick the part with the highest concavity (Both $C(M)$ and $C^*(M)$ can be used, but we find that using $C(M)$ produces better results).
We compute all visibility edges for that part and sample $k$ candidate cutting planes.
Each sampled plane is evaluated according to the objective function $Q_p(M,E)$ \eqref{eq:value}, and the plane with the highest value is selected.
The mesh is subsequently partitioned using this cutting plane.
In cases where the cut yields more than two disconnected components, we further separate these components using triangle-level connectivity.
The stopping criteria to terminate the decomposition process are: a) the target concavity threshold is reached; b) the prescribed number of parts has been obtained; or c) the current part contains no remaining visibility edges.

\section{Results and Experiments} 

\begin{table}
\centering
\begin{tabular}{@{}llcc@{}}
\toprule
Dataset & Method      & Concavity $\downarrow$ & Parts $\downarrow$ \\
\midrule
\multirow{4}{*}{V-HACD} & VisACD (Ours)        &     \textbf{0.043}      &     \textbf{28.4}  \\ 
& CoACD~\cite{wei2022approximate} &     0.048       &     31.6    \\ 
& V-HACD~\cite{mamou2016volumetric}      &     0.118       &     57.6    \\ 
& \citet{thul2018approximate}   &     0.069       &     34.4    \\ 
\midrule
\multirow{2}{*}{PartNet-Mobility} & VisACD (Ours) & 0.046 & \textbf{35.1} \\ 
& CoACD & 0.046 & 35.6\\
\midrule
\multirow{2}{*}{Objaverse} & VisACD (Ours) & 0.047 & \textbf{45.4} \\ 
& CoACD & 0.047 & 58.3 \\ 
\bottomrule
\end{tabular}
\caption{Quantitative comparisons. \visacd outperforms baselines, especially on Objaverse where meshes have more irregular orientation and geometric structure.}
\label{tab:evaluation_results}
\end{table}


\mypara{Task Definition.}
We decompose an input 3D mesh into a set of convex hulls using one of the ACD algorithms. We then report the quality of the decomposition through the number of convex hull parts, and concavity measuring distance from the input.

\mypara{Setup.}
We evaluate on V-HACD~\cite{mamou2016volumetric}, PartNet-Mobility~\cite{xiang2020sapien} and Objaverse~\cite{deitke2023objaverse}.
For Objaverse we use a randomly sampled subset of 1,000 meshes.
We compare results with CoACD~\cite{wei2022approximate}, V-HACD\cite{mamou2016volumetric} and \citet{thul2018approximate} (results reproduced from CoACD paper as source code is not publicly released).
We do not evaluate methods that produce decompositions with intersecting convex hulls.
We evaluate CoACD without the merging step, as merging produces intersecting convex hulls in 35\% of cases.
We do not use merging in our algorithm for the same reason.
We use $C(M)$ for the evaluation.
We also preprocess the meshes using SDF remeshing to make the meshes watertight, limit the number of vertices, and make vertex density more uniform.

\mypara{Discussion.}
We present the results in \Cref{tab:evaluation_results}.
Our method outperforms all baselines on all datasets while having lower computation time (16.97 seconds on average per PartNet-Mobility model vs 36.31 for CoACD). 
We attribute the smaller differences on PartNet-Mobility to the prevalence of highly regular objects such as tables, chairs, and bookcases, that can be decomposed into nearly perfect convex components using only axis-aligned cutting planes.
In contrast, the V-HACD and Objaverse datasets contain models with diverse, non-standard orientations and more organic shapes, including humans and animals, for which our method is well suited.








\begin{figure}
    \centering
    \includegraphics[width=0.67\linewidth]{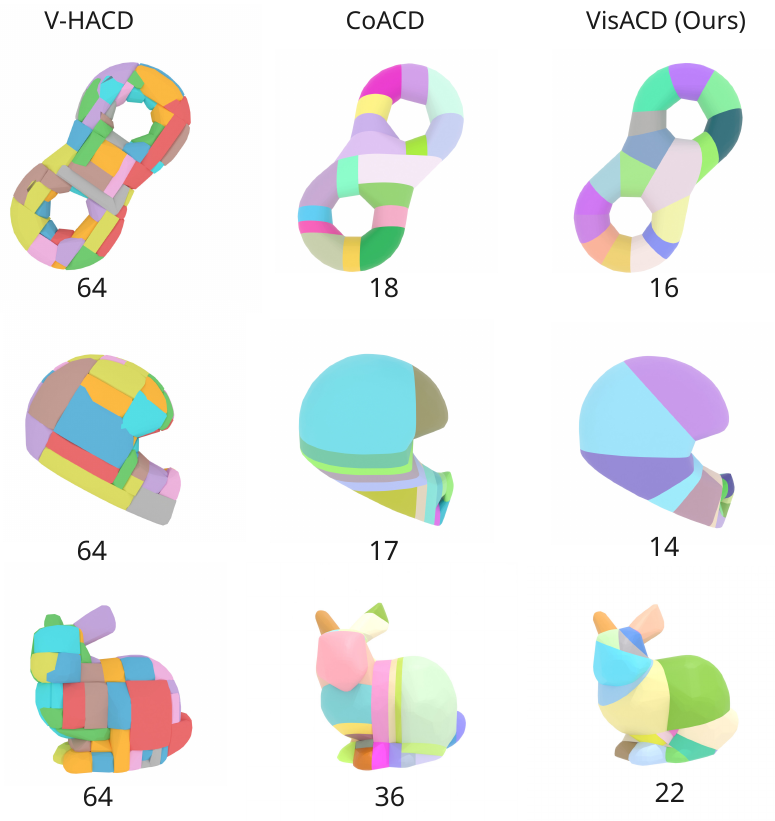}
    \caption{Qualitative Comparisons. We see that \visacd is not limited by axis-aligned planes and can produce accurate decompositions with fewer parts (indicated in numbers). }
    \label{fig:qualitative comparison}
\end{figure}

\section{Conclusion and Future Work}


We presented an approximate convex decomposition method that combines point visibility with cutting planes. It achieves a favorable trade-off between concavity and part count, while being significantly faster than prior work.
At the same time, there are limitations to address. The current pipeline relies on a greedy algorithm that can produce sub-optimal solutions. Our algorithm is also sensitive to the topology of the original mesh, and requires remeshing for optimal performance.
Future work may attempt to come up with an algorithm that considers the best possible decomposition not only on the current step, but also in the future (e.g. through Monte-Carlo Tree Search), while still utilizing the efficiencies of our approach.



\renewcommand*{\mkbibnamefamily}[1]{\textsc{#1}}
\renewcommand*{\mkbibnamegiven}[1]{\textsc{#1}}
\printbibliography

\end{document}